# Harnessing the Power of Decision Trees to Detect IoT Malware

Marwan Omar
Illinois Institute of Technology,
Chicago, USA

**Abstract:** Due to its simple installation and connectivity, the Internet of Things (IoT) is susceptible to malware attacks. Being able to operate autonomously. As IoT devices have become more prevalent, they have become the most tempting targets for malware. Weak, guessable, or hard-coded passwords, and a lack of security measures contribute to these vulnerabilities along with insecure network connections and outdated update procedures. To understand IoT malware, current methods and analysis ,using static methods, are ineffective. The field of deep learning has made great strides in recent years due to their tremendous data mining, learning, and expression capabilities, cybersecurity has enjoyed tremendous growth in recent years. As a result, malware analysts will not have to spend as much time analyzing malware. In this paper, we propose a novel detection and analysis method that harnesses the power and simplicity of decision trees. The experiments are conducted using a real word dataset, MaleVis which is a publicly available dataset. Based on the results, we show that our proposed approach outperforms existing state-of-the-art solutions in that it achieves 97.23% precision and 95.89% recall in terms of detection and classification. A specificity of 96.58%, F1-score of 96.40%, an accuracy of 96.43.

## 1. Introduction

An intrusion is an attempt to compromise security objectives by infecting a system. Therefore, many tools and methods have been developed to protect networks and systems from intrusion, such as detection systems [7–9]. Thus, intrusion detection consists of techniques that classify data activity into normal and intrusive [6, 8] to detect undesirable activity. An intrusion detection system detects and stops intruders from entering a monitored network outside or inside. Two methods of detection are generally used for this purpose. For example, misuse detection detects intrusion by using a known attack signature. Another detection method is anomaly detection, which is based on deviations from a normal model [1, 8, and 10]. By combining misuse detection with anomaly detection, hybrid detection approaches are aimed at increasing the detection rate and accuracy of IDS [9, 11, and 12].

These IDS are efficient, but they have a lot of limitations, like real-time detection, alarm generation, and data accuracy, which can lead to less successful detections [6, 8]. Because of this, intrusion detection is still an important and dynamic research field. We've been integrating ML methods to improve intrusion detection and strengthen computer security. Several papers have explored using machine learning to enhance data quality and training to improve intrusion detection performance [13–20]. There are many issues in which decision trees have been used for classification. You test each feature individually. After each branch is split, a single classification is assigned to it [21, 22]. It represents the training set better than a decision tree. It can predict their values because it includes the values of instances outside the training set. There is a guarantee that the decision tree constructed by ID3 and C4.5 will correspond to the data provided by both of these well-known algorithms.

Meanwhile, the data are not always gathered in a structured manner. Unstructured data must be preprocessed before they can be analyzed. Additionally, selecting relevant features is an important step aimed at reducing the computational costs of modelling and improving the predictive model's performance [13, 24].

An approach based on decision trees for network intrusion detection and making accurate decisions is proposed in this paper. The quality of the data was improved through the use of feature engineering. We have validated two major contributions in this study. The first step is to improve data quality by applying the entropy decision method. We also developed a classifier model based on decision tree algorithms to detect network intrusions effectively. Please find the following section for more information.

**Related Works**

Over the last decade, some intrusion detection techniques have been adopted to achieve computer security objectives. The goal of intrusion detection research is to increase the effectiveness and capability of IDS through automatic responses. ML techniques are becoming more widely used in intrusion detection [13–20]. As such, intrusion detection using ML is a classification project involving using labelled data to build classifiers that can determine the difference between normal and abnormal activity [11, 16, 21, 27, and 28]. They allow an effective classifier to be trained and built using relevant data [13, 17, 23, 25, 35, and 36].

An anomaly-based intrusion detection system based on fuzzy SOM method was proposed by Karami [37] in 2018. A model for intrusion detection combining NB and DL technique was proposed by Tabash et al. [26] in 2020. A genetic algorithm was used in the model for feature selection. The detection model proposed by Ghazali et al. [27] for intrusive communication was published in 2015. These methods include Simple Cart, NB, BFTree, PART, and Ridor. For network intrusion detection, It achieved an accuracy of 89.24% using NSL-KDD dataset.

Using the random forest algorithm, Hadi [29] proposed a model for selecting significant features in 2018. NSL-KDD was used to evaluate this model. According to this model are all within the CPI range. As a method to improve data quality, Gu et al. [17] proposed ensemble SVM-based intrusion detection with LMDRT transformations in 2019. Results show that the class performance on CICIDS2017 dataset is ACC 93.64%, DR 97.56%, and FAR 20.28%. A double PSO metaheuristic DL model was developed by Elmasry et al. [32] in 2020 to detect network intrusions. ACC, DR, and FAR values are generated on the CICIDS2017 dataset. According to Prasard et al. [36], a new IDS was proposed in 2019 that uses a probabilistic method in order to extract significant features from a subset of features.

A state-of-the-art literature review demonstrates that the learning methods and the data quality are important factors determining the robustness of IDS [6, 17, 26–29, 32, 36, 37].

Malware attacks can be detected and prevented using several methods. Systems for detecting and preventing incidents by intruders (IDPS) use several techniques. The data mining approach includes signature-based technologies, stateful protocol analyses, behavioral analyses and anomaly-based techniques. Different detection approaches are used in these technologies. Signature-based detection is an effective method for detecting already known threats but is not very effective when detecting previously unknown risks.

Similarly, they cannot be used to detect disguised threats when using evasion. On the other hand, anomaly-based screening can detect an invasion without requiring signatures. An unknown intrusion can instead be spotted by observing similar behaviour in other intrusions. Malware occurrences are identified through the concept of modelling normality associated with anomaly-based detection. Observing different activities and characteristics results in the creation of these sketches. It can uncover previously undiscovered threats because anomaly-based detection looks for abnormal patterns.

1. Novel Network Intrusion Detection Approach

3.1 A Proposed Model. This model comprises three main components, as shown in Figure 1: data quality, the building of a classifier, and deployment of intrusion detection. Here are details on what each of these components includes.

Part 1: Process of Data Quality.

Among its primary responsibilities is the collection and preprocessing of data. As a result, the system follows a process that gathers and accumulates data from networks. Following data collection, the gathered network traffic is subjected to data preprocessing. Preprocessing treats the data type incompatibilities and ignores them. A further step involved sanitizing the data and saving the results. The data is further transformed, resulting in the finalized features of the network dataset. These features are chosen using the entropy decision procedure.

## Part 2: Classifier Construction.

The second part will be started after the first has been completed. As the title indicates, the goal of the second component is to create a classifier model. Input is the data that was transformed in the data quality process. There are two main phases of classifier construction: training and validation of the model. The first phase of our proposed method involves training a decision tree classifier with three portions of data. Our model is then validated using the rest of the data in the second phase.

Part 3: Defining network intrusion detection systems.

For effective IDS to be improved, actual testing is necessary. Therefore, we can test its ability to differentiate between normal and abnormal behaviours. Thus, the IDS can be made more accurate based on classification results. To detect anomalous behaviour in systems, various approaches have been utilised in the literature. Researchers are very interested in intrusion detection and prevention since there is a concern about protecting systems from intrusion while maintaining network security regulations. This section describes the various strategies employed in prior research for intrusion detection and prevention.

*3.2.* Description of Proposed Solution. We have explained above that our approach begins with collecting and transforming data, followed by selecting features based on the needs of analysis and detection. Creating an intrusion detection model that is accurate requires good quality data. Thus, this step focuses on preparing data for analysis and making accurately informed decisions. To obtain a good training set, we first perform feature extraction using entropy decision on original traffic inside network traffic. It is a crucial step aimed at increasing the accuracy of our methodology. Besides reducing training complexity by analyzing less data, the goal is to achieve a great model that performs well in accuracy, detection rate, and real-time detection. Before analyzing network traffic, we apply specific preprocessing to collected data. It is then normalized. This process reduces the number of features in collected network data.

To improve our approach accuracy, we implement proposed data quality techniques to transform the data. Therefore, an effective intrusion detection model can be trained and validated by using the decision tree to make timely decisions. Furthermore, intrusion detection is

Cybersecurity is increasingly becoming an established topic with the emphasis on securing sensitive data stored in computer systems and networks. By blocking unauthorized access to the system, attackers can't obtain, corrupt, damage, or destroy this information. A computer intrusion detection system identifies any breach of security in a computer network by revealing activity in the network. Detecting potential malware and its ability to cause harm is the purpose of an intrusion detection system (IDS). We must recover from malware and detect it at this point. An intrusion prevention system (IPS) detects malware in operations and also tries to prevent the occurrence of potential incidents. Although the internet has changed a lot of things, it has also presented a number of insecurity areas that need to be addressed in a secure environment to be maintained.

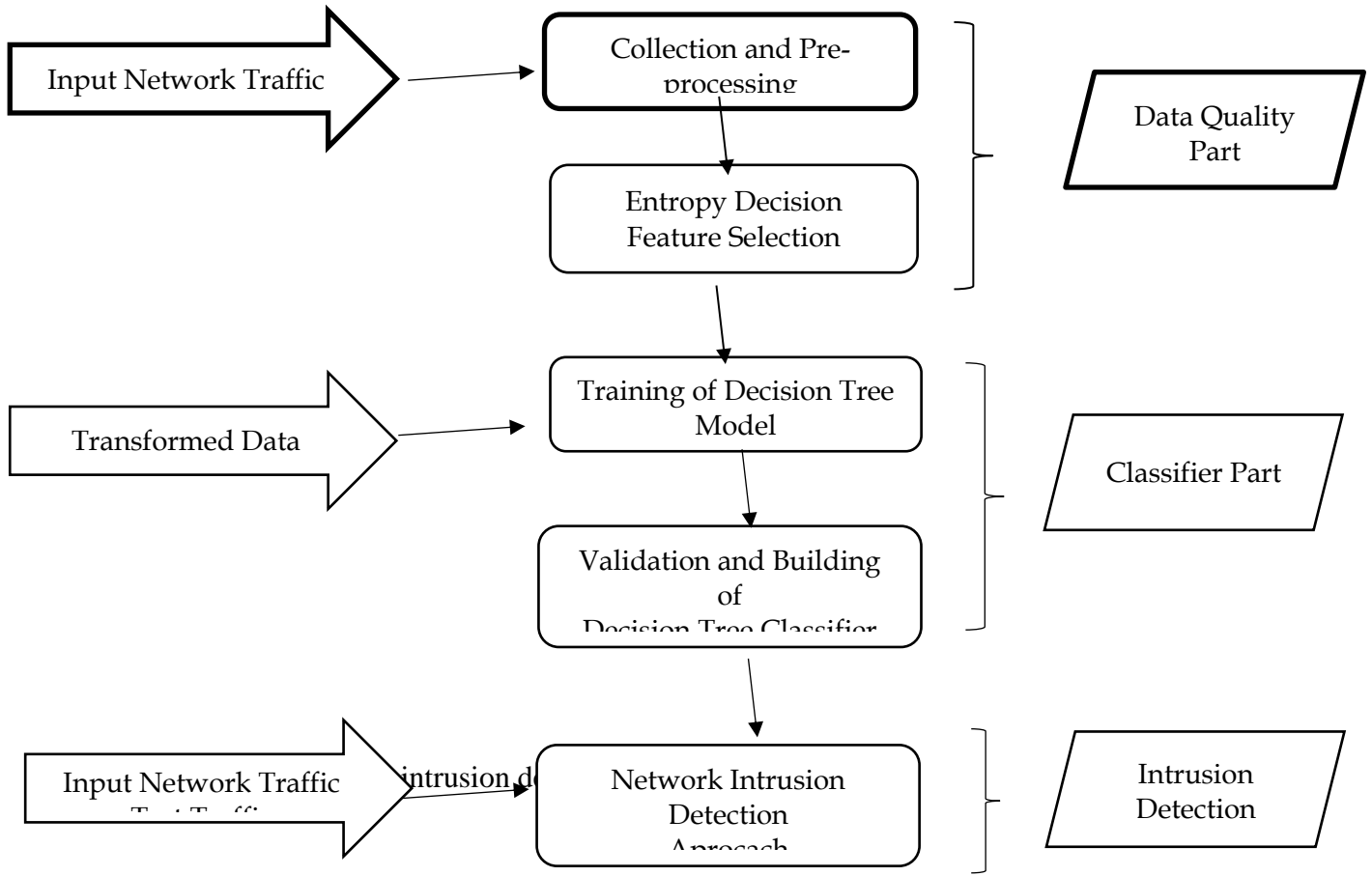

**Experimental Results and Discussion**

Validating intrusion detection programs rely heavily on the evaluation of datasets. As a result, to evaluate an IDS using ML techniques, many available and appropriate datasets are available. We have chosen two types of datasets to evaluate and validate our approach, including NSL-KDD and CICIDS2017. The NSLKDD dataset is very practical because of its novelty and the sheer number of examples it contains. CICIDS2017 was created from a dataset created by Canadian Institute for Cyber Security. Presented here is an effective intrusion detection dataset that overcomes the limitations of the actual dataset.

**Table 1.** Hyperparameters deployed for model best performance.

| Hyperparameters | Values |
|---|---|
| Batch size | 64 |
| Number of Epochs | 25 |
| Learning rate | 1e-3 |
| Optimization Algorithm | Adam |
| Loss function | Cross-entropy |

**Table 2.** Shows the distribution of our Malevis dataset.

| Category | Type | Class | Samples | Total |
|---|---|---|---|---|
| Benign | - | Normal | 1832 | 1832 |
| Malware | Adware | Adposhel | 494 | 2983 |
| | | Amonetize | 497 | |
| | | BrowseFox | 493 | |
| | | InstallCore | 500 | |
| | | MultiPlug | 499 | |
| | | Neoreklami | 500 | |
| | Trojan | Agent | 470 | 4440 |
| | | Dinwod | 499 | |
| | | Elex | 500 | |
| | | HackKMS | 499 | |
| | | Injector | 495 | |
| | | Regrun | 485 | |
| | | Snarasite | 500 | |
| | | VBKrypt | 496 | |
| | | Vilsel | 496 | |
| | Virus | Neshta | 497 | 1997 |
| | | Sality | 499 | |
| | | Expiro | 501 | |
| | | VBA | 500 | |
| | Worm | Allaple | 478 | 1974 |
| | | Autorun | 496 | |
| | | Fasong | 500 | |
| | | Hlux | 500 | |
| | Backdoor | Androm | 500 | 1000 |
| | | Stantinko | 500 | |

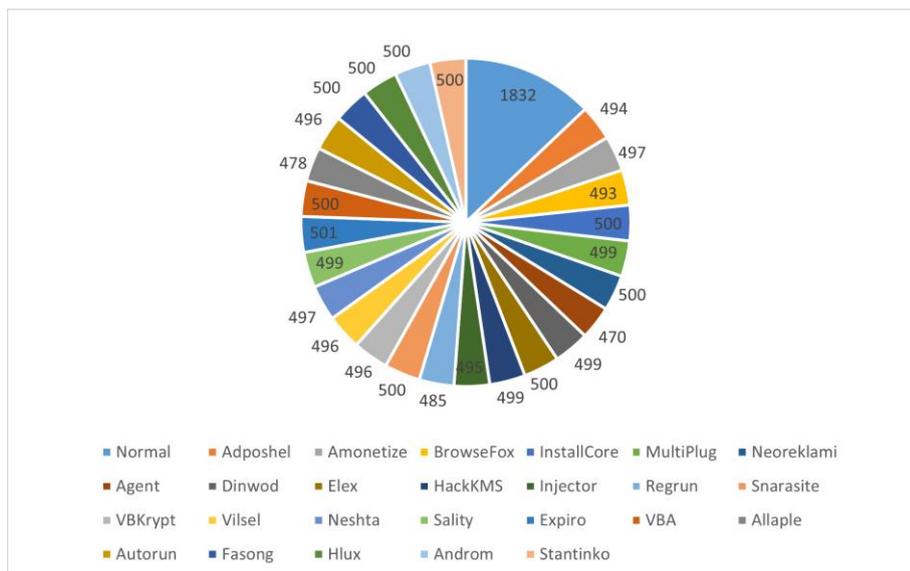

Figure 2- shows the distribution of data samples across different classes of the dataset.

Table 3- shows the performance of our decision tree based approach

| Model | Spec (%) | F1-Score (%) | Acc (%) | Loss | APT per Malware Classification (ms) |
|---|---|---|---|---|---|
| ResNet18 | 94.14 | 96.39 | 95.03 | 0.181 | 146 |
| Our approach | 98.69 | 98.76 | 97.67 | 0.086 | 97 |
| DenseNet161 | 97.97 | 94.67 | 96.66 | 0.156 | 480 |

We need intrusion detection and prevention systems to protect our systems and networks daily from attacks and intrusions. The only way to collect data is through literature reviews. A discussion will be made on the various techniques, their potential for coping with such attacks, and their limitations. Different approaches have been used in the literature to detect abnormal behaviour in systems. Researchers have developed an interest in intrusion detection and prevention since it has become important to protect systems from intrusions by preserving network security protocols. There is a discussion here of the various mechanisms used in previous research for intrusion detection and prevention. To develop a model for detecting abuse, a decision tree was first used. Using this structure, drill information was disintegrated into more simplistic subsets. A vector machine was also operating simultaneously to support the development of anomaly detection hypotheses in every area. To enhance the model's functionality, it can use the information of malware identified while creating contours for standard behaviors.

In the detection of system intrusions, IDPS methods present several benefits. At the same time, they present several limitations. Unidentified novel exposures cannot be detected by signature detection alone; therefore, constant updating of the database is essential. The upkeep of IDS is also time-consuming, making it difficult to update new attacks. Identifiable device protection systems that use anomaly-based detection are likely to suffer from deceptive alarms at a considerable rate. During the profile construction and training phases, the system is usually not monitored, making the activities skipped during those phases illegitimate. In addition, the profile of the typical behaviour contours file is continuously updated due to the user's behavior changing most of the time. During the phase of construction and training, the system is also unable to detect any anomalies. Several methods are now available to detect possible malware in Internet-connected devices; therefore, the safety of information systems is becoming a major concern. IDPs can be built using multiple mechanisms, including signature-based exposure and anomaly detection. Methods like classification and clustering are used in these methods to detect malware. Hackers are constantly developing new methods of intrusion. Therefore, IDPS requires a great deal of improvement to enhance its intrusion detection and prevention level due to its limitations.

Anomaly-based screening, on the other hand, does not require signatures to detect a system intrusion. Instead, it can detect an unknown incursion by observing comparable intrusion patterns. The concept of modelling normality is used in the anomaly-based detection strategy to discover malware instances. An outline represents the usual behaviour of items like network users, applications, connections, and hosts in an intrusion detection system that uses anomaly-based exposure. After some time of witnessing various activities and traits, these sketches come into play. Anomaly-based detection has the benefit of allowing previously unknown dangers to be discovered.

**Conclusion and Future Works**

Monitoring systems and data for security is a key step in the intrusion detection process. The paper presents a method for detecting network intrusions using a decision-tree-based classifier and engineering features. To increase the detection rate and accuracy of IDS, a preprocessing phase is being set up based on the heterogeneity of the data. Consequently, the novel model proposed for detecting network intrusions has many advantages and is highly accurate. We will implement other efficient machine learning techniques such as deep learning in various parts of our method in the future to improve the detection rate and accuracy